\documentclass[a4paper,twoside]{article}

\usepackage{epsfig}
\usepackage{subcaption}
\usepackage{calc}
\usepackage{amssymb}
\usepackage{amstext}
\usepackage{amsmath}
\usepackage{amsthm}
\usepackage{multicol}
\usepackage{pslatex}
\usepackage{apalike}
\usepackage{booktabs}
\usepackage{algorithm2e}
\usepackage[bottom]{footmisc}
\usepackage{SCITEPRESS}     

\begin{document}

\title{Emotional Modulation in Swarm Decision Dynamics}

\author{\authorname{David Freire-Obregón\orcidAuthor{0000-0003-2378-4277}}
\affiliation{SIANI, Universidad de Las Palmas de Gran Canaria, Las Palmas de Gran Canaria, Spain}
\email{david.freire@ulpgc.es}
}


\abstract{Collective decision-making in biological and human groups often emerges from simple interaction rules that amplify minor differences into consensus. The bee equation, developed initially to describe nest-site selection in honeybee swarms, captures this dynamic through recruitment and inhibition processes. Here, we extend the bee equation into an agent-based model in which emotional valence (positive–negative) and arousal (low–high) act as modulators of interaction rates, effectively altering the recruitment and cross-inhibition parameters. Agents display simulated facial expressions mapped from their valence–arousal states, allowing the study of emotional contagion in consensus formation.
Three scenarios are explored: (1) the joint effect of valence and arousal on consensus outcomes and speed, (2) the role of arousal in breaking ties when valence is matched, and (3) the ``snowball'' effect in which consensus accelerates after surpassing intermediate support thresholds. Results show that emotional modulation can bias decision outcomes and alter convergence times by shifting effective recruitment and inhibition rates. At the same time, intrinsic non-linear amplification can produce decisive wins even in fully symmetric emotional conditions. 
These findings link classical swarm decision theory with affective and social modelling, highlighting how both emotional asymmetries and structural tipping points shape collective outcomes. The proposed framework offers a flexible tool for studying the emotional dimensions of collective choice in both natural and artificial systems.
}

\onecolumn \maketitle \normalsize \setcounter{footnote}{0} \vfill

\section{Introduction}
\label{sec:introduction}

Collective decision-making is a fundamental process observed in both biological and human systems, from honeybee swarms selecting a new nest site to committees, crowds, and online communities reaching agreement on shared choices. Honeybee decision dynamics, in particular, have been effectively modeled using the ``bee equation'', which captures how recruitment and cross-inhibition between competing options can drive the emergence of consensus \cite{Seeley2011,Chase25}. This formulation has been widely adopted in studies of swarm intelligence and distributed coordination due to its explanatory power and simplicity.

In parallel, research in psychology and affective computing has demonstrated that social influence is not purely rational or informational: emotional cues, often conveyed through facial expressions, body language, or tone of voice, play a significant role in shaping preferences and accelerating or slowing agreement \cite{Hatfield1994,Freire25abm}. The dimensional model of affect, describing emotions along the axes of valence (positive–negative) and arousal (low–high), is widely used to represent and quantify emotional states in computational models \cite{Russell1980}. Empirical studies have shown that both the sign and intensity of emotional displays can modulate persuasiveness and trust, influencing how opinions spread in face-to-face and mediated communication.

While these two strands of research—swarm decision models and emotion-based influence—have each developed robust theoretical and empirical foundations, integration between them remains limited. Most applications of the bee equation have focused on recruitment signals without explicitly modeling the emotional state of individuals or the contagion of affect. Conversely, models of emotional contagion in social networks rarely incorporate the inhibitory dynamics and feedback loops characteristic of swarm decision-making \cite{vanHaeringen2021}. This gap limits understanding of how emotional variables interact with recruitment dynamics to determine not only the final decision outcome but also the temporal profile of consensus formation.

This study addresses this gap by extending the bee equation into an agent-based model in which each agent carries an evolving emotional state defined by valence and arousal. These variables are mapped to simulated facial expressions, which in turn modulate recruitment and inhibition rates in the decision process. Emotional contagion occurs between agents, allowing the spread of affective states to amplify or dampen support for each option. This framework enables the exploration of scenarios where emotional cues are the primary differentiator between otherwise identical choices.

Three experimental scenarios are investigated: (1) the joint effect of valence and arousal on consensus probability and time to agreement, (2) the role of arousal in breaking ties when valence is matched, and (3) the ``snowball'' effect in which consensus accelerates after intermediate support thresholds are surpassed. Results indicate that valence–arousal dynamics can significantly bias outcomes even when options are objectively equivalent, and that rapid early growth in support strongly predicts eventual victory. By integrating emotional contagion into a well-established swarm decision model, this work bridges swarm intelligence, affective computing, and social simulation, offering a flexible framework for examining the emotional dimensions of collective choice.

\begin{figure}[ht]
    \centering
    \includegraphics[width=\columnwidth]{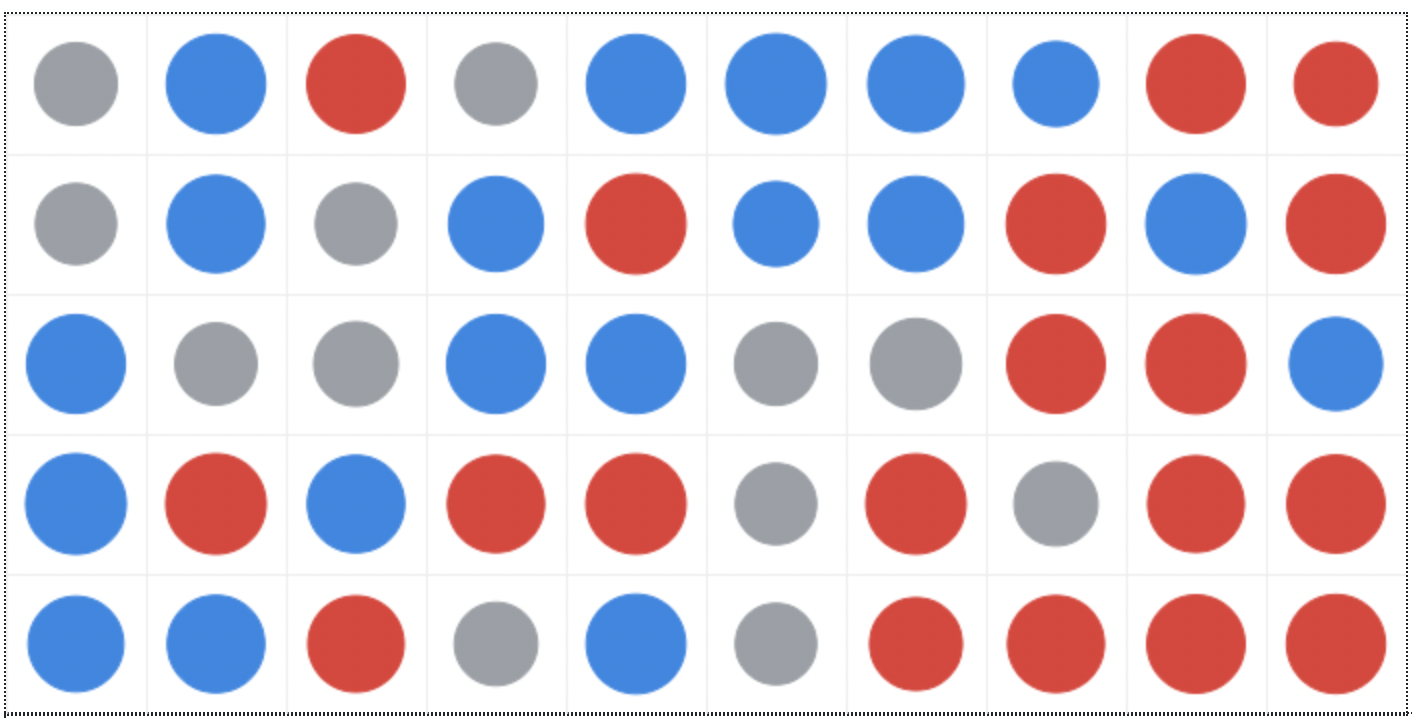}\\[0.5em]
    \includegraphics[width=\columnwidth]{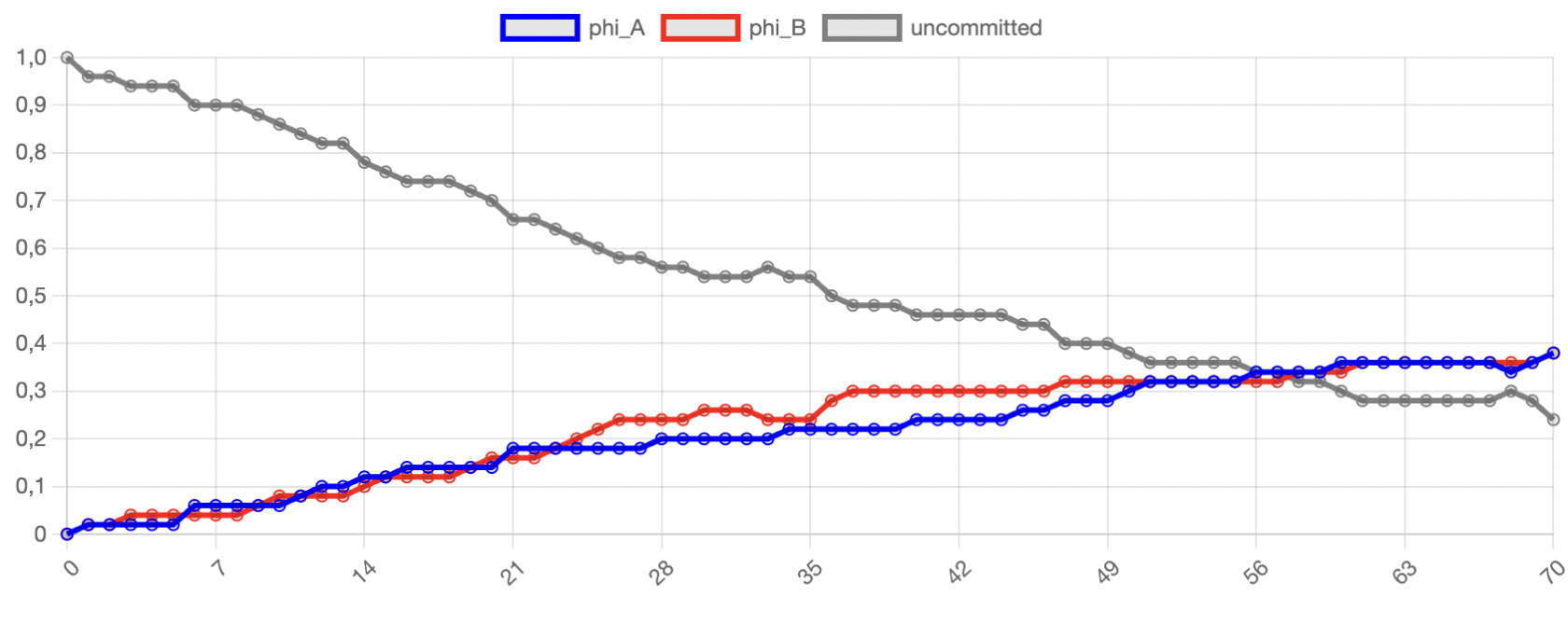}
    \caption{Agent-based simulation of the bee decision-making model. 
    \textbf{Top:} spatial distribution of agents on the grid, with blue representing commitment to option A, red to option B, and grey uncommitted individuals. 
    \textbf{Bottom:} temporal evolution of the proportion of agents in each state ($\varphi_A$ in blue, $\varphi_B$ in red, and uncommitted $u$ in grey). 
    The dynamics illustrate the gradual recruitment from the uncommitted pool towards both options and the role of cross-inhibition in shaping the relative growth of A and B.}
    \label{fig:bee_model_simulation}
\end{figure}

\section{Related Work}

\textbf{Swarm Decision Models and the Bee Equation.}  
The bee equation was initially formulated to describe nest-site selection in honeybee swarms \cite{LAOMETTACHIT201521}, where scout bees recruit others to inspect potential sites while also delivering inhibitory stop signals to competing recruiters \cite{Seeley2011,BERNARDI2018158}. This positive–negative feedback mechanism enables the swarm to reach a collective decision without centralized control. Variants of the bee equation have been developed to analyze decision dynamics under noise, time constraints, or network structure \cite{Chase25}. Beyond biological systems, the bee equation has been applied to swarm robotics, collective search, and distributed sensor networks, where robustness and scalability are critical. An influential formalisation of these dynamics was provided by Pais et al.~\cite{pais2013mechanism}, who extended the bee equation into a value-sensitive decision framework that generalises recruitment and inhibition parameters for asymmetric options.

\textbf{Emotional Contagion and Valence–Arousal Modeling.}  
In human groups, decision-making is strongly influenced by emotional states that spread through observation of facial expressions, tone of voice, and other social cues \cite{SHI2021769}. This process, known as emotional contagion, is well documented in psychology and behavioral science \cite{Hatfield1994}. Emotions are often modeled in a two-dimensional affective space, with valence representing pleasantness and arousal representing activation \cite{Russell1980}. The valence–arousal model has proven helpful in computational studies of crowd dynamics, online social networks, and human–computer interaction, allowing a compact but expressive representation of emotional states.

\textbf{Agent-Based Models Combining Social and Emotional Dynamics.}  
Agent-based modeling (ABM) has been widely used to study how local interactions give rise to emergent patterns in social systems \cite{TEMPLETON2024106520}. In the context of emotional contagion, ABMs can simulate heterogeneous populations where individuals update both cognitive and affective states based on interactions \cite{3577589}. For example, van Haeringen et al. \cite{vanHaeringen2021} integrated valence–arousal dynamics into an ABM to simulate emotion spread in crowds, showing how different emotional profiles affect global behavior. However, most ABMs of emotion propagation omit the explicit recruitment and inhibition terms characteristic of swarm models like the bee equation, limiting their ability to capture specific competitive decision scenarios.

\textbf{Research Gap.}  
While swarm decision models offer a principled framework for consensus formation and emotional contagion models capture affective influences on group dynamics, their integration remains scarce. There is little work examining how emotional valence and arousal interact with the recruitment–inhibition dynamics of the bee equation to influence both the outcome and the timing of collective decisions. Addressing this gap can provide new insights into the role of emotion in distributed coordination, with implications for both natural and artificial systems.

\section{Methodology}

The model is implemented as an agent-based system inspired by the bee equation for collective decision-making \cite{Seeley2011,Chase25}, extended to include emotional valence–arousal dynamics \cite{Russell1980,vanHaeringen2021}. The environment is represented as a two-dimensional grid, where each agent occupies a discrete cell and can interact with neighboring agents within a Moore neighborhood (eight adjacent cells). Periodic boundary conditions are applied to avoid edge effects.

\subsection{Agent States and Emotional Representation}

Each agent $i$ has three main state variables: (1) its current decision state $d_i \in \{A,B,U\}$, where $A$ and $B$ denote commitment to one of the two options and $U$ denotes an uncommitted state; (2) its emotional valence $v_i \in [-1,1]$, with positive values indicating pleasantness and negative values indicating unpleasantness; and (3) its arousal $a_i \in [0,1]$, representing the activation level. Valence and arousal jointly define the agent's affective state in a two-dimensional space, which is mapped to a simulated facial expression used in visualizations.

\subsection{Bee Equation with Emotional Modulation}

The original ``bee'' decision-making model can be obtained as a reduced, symmetric case of the more general value-sensitive decision framework proposed by Pais et al. for collective choices in honeybee swarms~\cite{pais2013mechanism}. 

\paragraph{Population fractions.}
Let $\varphi_A(t)=N_A(t)/N$ and $\varphi_B(t)=N_B(t)/N$ denote the fractions of agents committed to options $A$ and $B$ at time $t$, and let $u(t)=1-\varphi_A(t)-\varphi_B(t)$ be the uncommitted fraction. We use the $\varphi$ notation throughout.

In this simplified formulation, the change in the number of committed individuals is governed by two processes: 
(i) \emph{social recruitment}, where committed individuals persuade uncommitted ones to join their option at a constant rate $r$, and 
(ii) \emph{cross-inhibition}, where committed individuals of one option send stop signals to those of the rival option at a constant rate $\sigma$. 
Under the assumptions of equal recruitment and inhibition rates for both options, no spontaneous abandonment, and ignoring stochastic fluctuations, the dynamics reduce to:
\begin{equation}
\frac{d\varphi_A}{dt} = r\,\varphi_A \, u - \sigma\,\varphi_A\,\varphi_B,
\end{equation}
\begin{equation}
\frac{d\varphi_B}{dt} = r\,\varphi_B \, u - \sigma\,\varphi_A\,\varphi_B,
\end{equation}
where $u$ is the proportion of uncommitted agents. 
In the classical form, $r$ and $\sigma$ are positive constants.

In this work, both recruitment and inhibition rates are modulated by emotional states. Specifically, for an interaction between recruiter $i$ and target $j$, the recruitment rate is defined as:

\begin{equation}
r_{ij} = r_0 \left[ 1 + \alpha_v \, v_i + \alpha_a \, a_i \right]
\end{equation}

The inhibition rate between a committed agent $i$ (the inhibitor) and a committed agent $j$ of the opposing option (the target) is defined as:
\begin{equation}
\sigma_{ij} = \sigma_0 \left[ 1 - \beta_v \, v_j + \beta_a \, a_i \right],
\end{equation}
where $\sigma_0$ is the baseline inhibition rate, $\beta_v$ and $\beta_a$ are sensitivity parameters, $v_j$ is the valence of the target, and $a_i$ is the arousal of the inhibitor. 
The term $-\beta_v v_j$ captures the idea that the target's positive valence reduces its susceptibility to inhibition, whereas negative valence increases it. 
For example, if $v_j > 0$, the subtraction lowers $\sigma_{ij}$, making it harder for the inhibitor to convince the target to abandon its choice; if $v_j < 0$, the subtraction becomes an addition, increasing $\sigma_{ij}$ and making inhibition more likely. 
The term $+\beta_a a_i$ models the effect of the inhibitor's arousal: higher arousal increases the effectiveness of the inhibition attempt, regardless of valence.

\subsection{Emotional Contagion}

When two agents interact, the emotional state of the target $j$ is updated according to a contagion rule:

\begin{equation}
v_j(t+1) = v_j(t) + \gamma_v \, \left[ v_i(t) - v_j(t) \right]
\end{equation}
\begin{equation}
a_j(t+1) = a_j(t) + \gamma_a \, \left[ a_i(t) - a_j(t) \right]
\end{equation}

where $\gamma_v$ and $\gamma_a$ are contagion rates for valence and arousal, respectively. This formulation implements a simple linear convergence towards the emotional state of the influencer.

\subsection{Agent Decision Rules}

At each discrete time step:
\begin{enumerate}
    \item Agents are updated in random order to avoid synchronous artifacts.
    \item A committed agent attempts to recruit an uncommitted neighbor with probability $r_{ij}$.
    \item A committed agent may attempt to inhibit a neighbor committed to the opposite option with probability $\sigma_{ij}$.
    \item Emotional contagion occurs whenever two agents interact, regardless of recruitment or inhibition outcome.
    \item Uncommitted agents adopt the option of the recruiter if recruited; committed agents may revert to the uncommitted state if inhibited.
\end{enumerate}

\subsection{Initialization and Parameters}

Simulations start with a specified proportion of agents committed to each option, with the rest uncommitted. Initial valence and arousal distributions are scenario-dependent: for example, in scenario 1, agents committed to $A$ may start with high positive valence and moderate arousal, while those committed to $B$ may start with neutral valence and high arousal. Model parameters $(r_0, \sigma_0, \alpha_v, \alpha_a, \beta_v, \beta_a, \gamma_v, \gamma_a)$ are tuned to explore different sensitivities to emotional modulation.

\subsection{Stopping Conditions and Metrics}

The simulation terminates when consensus is reached (all participants agree on one option) or after a maximum number of steps. Key metrics include:
\begin{itemize}
    \item \textit{Consensus time:} number of steps until all agents commit to the same option ($\varphi_A=1$ or $\varphi_B=1$).
    \item \textit{Win rate:} fraction of simulations where a given option prevails.
    \item \textit{Half-life:} time to reach 50\% commitment to the winning option ($\varphi_X=0.5$).
    \item \textit{Emotional trajectory:} average valence and arousal over time for agents in each state.
\end{itemize}

\section{Experimental Setup}

Three experimental scenarios are considered in order to disentangle the effects of emotional states on collective decision dynamics and to explore specific hypotheses derived from the literature. The first, \textit{Scenario 1: Joint Valence–Arousal Influence}, assigns different initial valence and arousal levels to agents committed to options $A$ and $B$. This configuration makes it possible to evaluate the combined effect of these two emotional dimensions on both the probability of a given option prevailing and the time required to reach consensus. The choice of varying both variables simultaneously responds to the need to examine potential interaction effects, as previous studies often address valence and arousal separately.

In \textit{Scenario 2: Arousal Tie-Break}, agents committed to $A$ and $B$ begin with identical initial valence but differ in arousal. This design isolates the specific contribution of arousal to decision outcomes, particularly in situations where neither option has an initial emotional advantage in terms of pleasantness or positivity. This situation is driven by empirical evidence indicating greater persuasive ability and faster decision shift attendant on higher arousal, and which may resolve otherwise frozen deadlocks in group interaction.

Finally, \textit{Scenario 3: Snowball Effect} starts with both options perfectly balanced in number and emotional state. The focus here is to investigate whether surpassing intermediate thresholds of support can trigger a self-reinforcing recruitment process that accelerates consensus formation. This is relevant for understanding tipping points in collective decisions, where a slight numerical advantage might cascade into rapid dominance, a phenomenon observed in both biological swarms and human social systems.

The selection of these scenarios provides complementary perspectives: the first explores interaction effects between valence and arousal, the second isolates a single emotional driver under controlled conditions, and the third examines non-linear amplification mechanisms in balanced competitions. Together, they form a coherent set of experiments that test how emotional modulation within the bee-equation framework can alter both the trajectory and outcome of collective decision-making.

\textbf{Implementation details}. The model is defined by a set of parameters that govern interaction dynamics and emotional contagion rates. The baseline recruitment rate is set to $r_0 = 0.02$, controlling how likely a committed agent is to recruit an uncommitted neighbor in the absence of emotional influence. The baseline inhibition rate is $\sigma_0 = 0.02$, determining the likelihood of a committed agent causing an opponent to revert to the uncommitted state. 

Emotional modulation of recruitment is governed by $\alpha_v = 0.5$ and $\alpha_a = 0.5$, which quantify the influence of valence and arousal, respectively, on the probability of successful recruitment. Similarly, $\beta_v = 0.5$ and $\beta_a = 0.5$ control how valence and arousal modulate the probability of successful inhibition. Positive valence generally increases recruitment effectiveness and reduces susceptibility to inhibition, whereas high arousal increases both recruitment and inhibition rates.

The parameters for emotional contagion are $\gamma_v = 0.1$ and $\gamma_a = 0.1$, which determine the rate at which the emotional state of a target agent shifts toward that of an interacting agent during social contact.

The spatial environment is a $20 \times 20$ grid (400 agents) with periodic boundaries, where a single agent occupies each cell. The initial distribution of commitment states and the emotional values assigned to agents depend on the experimental scenario. Uncommitted agents typically begin with $v = 0$ and $a = 0.5$, representing neutral valence and moderate arousal. Committed agents receive scenario-specific initial valence and arousal values, which may be fixed or drawn from truncated normal distributions centered at the scenario's specified means with a standard deviation of $0.05$ to introduce heterogeneity. Parameter values may vary between scenarios to explore different behavioral regimes.

\textbf{Simulation Protocol}.
Each experimental condition is replicated 200 times to account for stochastic variability. The maximum simulation length is 500 steps, with early termination if consensus is reached. At each simulation step, the fraction of agents committed to each option ($\varphi_A$ and $\varphi_B$), the fraction uncommitted ($u$), and the mean valence and arousal of each subgroup are recorded. After the simulations conclude, additional metrics are computed, including the \textit{Consensus Time}, defined as the number of steps required for all agents to support the same option; the \textit{Win Rate}, given by the fraction of runs in which a particular option prevails; the \textit{Half-Life}, corresponding to the number of steps needed to reach 50\% commitment to the eventual winning option; and the \textit{Emotional Trajectory}, which represents the temporal evolution of the average valence and arousal across the population.

\section{Evaluation of Simulation Scenarios}

We systematically evaluated the three experimental scenarios described in the previous section to assess how emotional valence and arousal influence the dynamics of collective decision-making in the bee-equation framework. Each scenario was simulated over multiple independent runs to ensure statistical robustness.  

The following paragraphs present the results for Scenarios~1, 2, and 3, emphasizing key trends and deviations from the baseline symmetric case.

\begin{table}[t]
\centering
\caption{Scenario~1: Joint Valence--Arousal Influence}
\label{tab:sc1}
\begin{tabular}{rrr r r}
\toprule
$a_A$ & $v_A$ & Win(A) & $\overline{t}_{\mathrm{cons}}$ & AUC$_A-$AUC$_B$\\
\midrule
0.20 & 0.20 & 0.40 & 315.1 & -0.125\\
0.20 & 0.50 & 0.50 & 303.7 &  0.029\\
0.20 & 0.80 & 0.65 & 305.2 &  0.139\\
0.20 & 1.00 & 0.50 & 338.6 & -0.018\\
0.50 & 0.20 & 0.65 & 318.8 &  0.068\\
0.50 & 0.50 & 0.85 & 303.4 &  0.265\\
0.50 & 0.80 & 0.40 & 356.6 & -0.076\\
0.50 & 1.00 & 0.75 & 312.1 &  0.288\\
0.80 & 0.20 & 0.70 & 336.1 &  0.148\\
0.80 & 0.50 & 0.55 & 330.0 &  0.082\\
0.80 & 0.80 & 0.70 & 318.5 &  0.190\\
0.80 & 1.00 & 0.75 & 303.2 &  0.283\\
1.00 & 0.20 & 0.65 & 275.2 &  0.166\\
1.00 & 0.50 & 0.55 & 304.5 &  0.057\\
1.00 & 0.80 & 0.80 & 311.0 &  0.313\\
1.00 & 1.00 & 0.80 & 273.1 &  0.312\\
\bottomrule
\end{tabular}
\end{table}

\noindent

\textit{Scenario 1}. Table~\ref{tab:sc1} summarizes the effect of initial valence ($v_A$) and arousal ($a_A$) on the outcome of this scenario. The table reports, for each combination of initial arousal ($a_A$) and initial valence ($v_A$) assigned to agents committed to option~A, the proportion of simulation runs in which option~A prevailed (Win(A)), the mean time to consensus in simulation steps ($\overline{t}_{\mathrm{cons}}$), and the difference between the areas under the commitment curves for options~A and~B (AUC$_A-$AUC$_B$), which quantifies the relative dominance of~A throughout the decision process.

Two robust trends emerge. First, higher $a_A$ generally increases the win rate of option~A, especially when coupled with moderate-to-high $v_A$, indicating that emotional intensity amplifies the persuasive impact of positive valence. 
Second, reductions in the mean time to consensus are more pronounced in the same region of the $(v_A,a_A)$ plane, suggesting faster recruitment dynamics when agents display both pleasant and energetic expressions. 
The AUC difference (AUC$_A-$AUC$_B$) aligns with these patterns, showing sustained dominance of A over the trajectory when $(v_A,a_A)$ are jointly elevated. 
Taken together, these results support an interaction between valence and arousal: either dimension alone has a limited effect, but their combination consistently shifts both the probability of prevailing and the speed of convergence.

\begin{figure*}[t]
    \centering
    \includegraphics[width=0.6\textwidth]{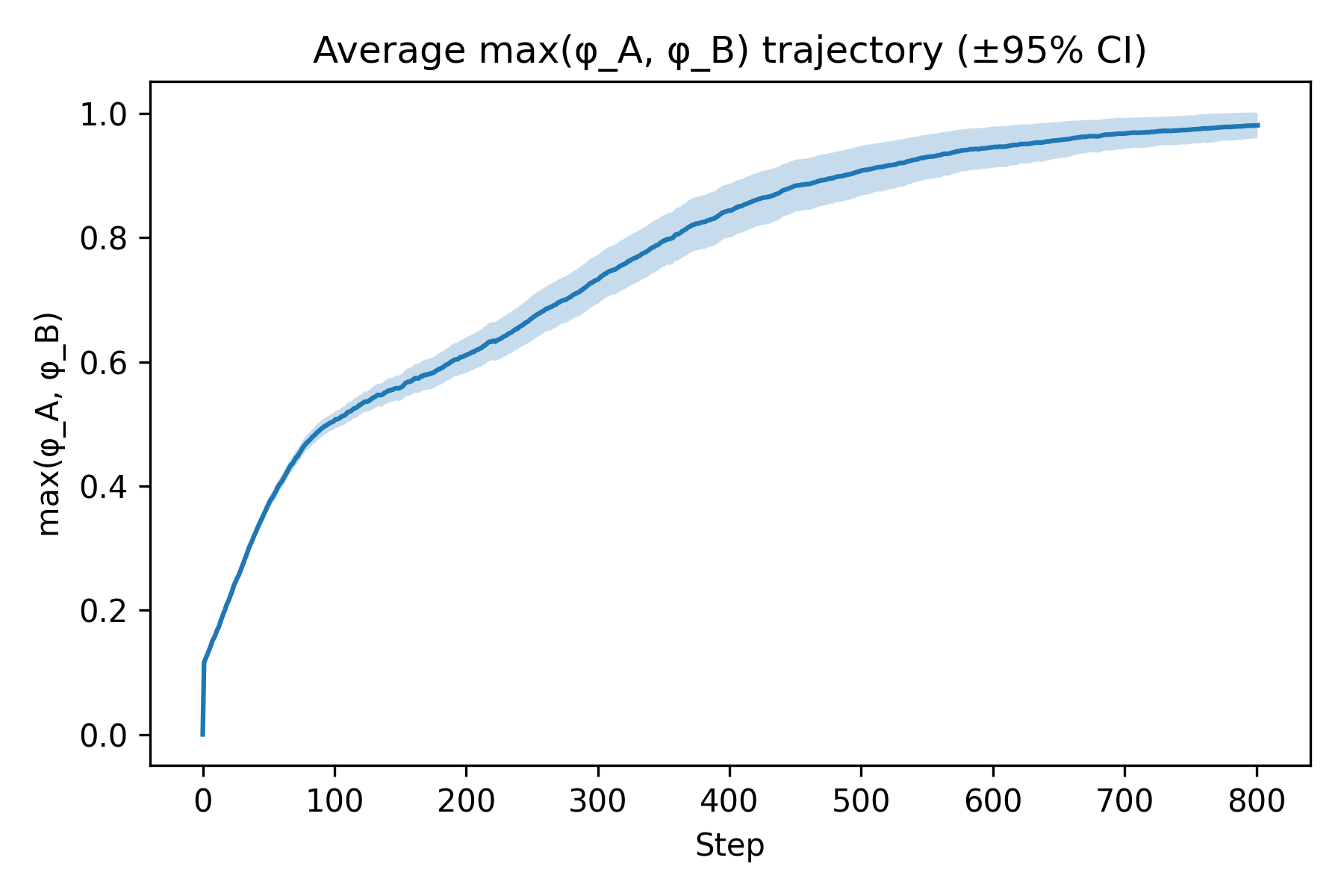}
    \caption{Scenario~3: Snowball Effect. 
    The plot shows the average trajectory of $\max(\varphi_A, \varphi_B)$, where $\varphi_X$ denotes the proportion of agents committed to option $X$, across simulation steps. 
    The shaded area indicates the $95\%$ confidence interval. 
    Starting from a perfectly balanced configuration, the system exhibits a gradual increase in the maximum commitment level until one option approaches complete dominance. 
    This reflects a self-reinforcing recruitment dynamic: once an intermediate threshold of support is surpassed, the leading option continues to attract uncommitted agents at an accelerating pace.}
    \label{fig:sc5}
\end{figure*}
\noindent

\begin{table}[t]
\centering
\caption{Scenario~2: Arousal Tie-Break.}
\label{tab:sc2}
\begin{tabular}{rrr r}
\toprule
$a_A$ & $v_A$ & Win(A) & $\overline{t}_{\mathrm{cons}}$\\
\midrule
0.20 & 0.20 & 0.80 & 335.6\\
0.20 & 0.50 & 0.60 & 329.6\\
0.20 & 0.80 & 0.75 & 308.5\\
0.20 & 1.00 & 0.60 & 297.7\\
0.50 & 0.20 & 0.50 & 296.1\\
0.50 & 0.50 & 0.65 & 284.2\\
0.50 & 0.80 & 0.65 & 358.4\\
0.50 & 1.00 & 0.75 & 287.4\\
0.80 & 0.20 & 0.60 & 261.6\\
0.80 & 0.50 & 0.65 & 337.1\\
0.80 & 0.80 & 0.65 & 322.8\\
0.80 & 1.00 & 0.65 & 325.1\\
1.00 & 0.20 & 0.55 & 307.7\\
1.00 & 0.50 & 0.70 & 325.0\\
1.00 & 0.80 & 0.70 & 320.1\\
1.00 & 1.00 & 0.65 & 330.3\\
\bottomrule
\end{tabular}
\end{table}

\noindent
Table~\ref{tab:sc2} shows the outcomes for scenario~2, where valence is fixed across both options and only arousal varies for agents initially committed to option~A. Results are shown for each combination of initial arousal ($a_A$) and valence ($v_A$) assigned to agents committed to option~A, with valence kept identical across both options to isolate the effect of arousal. 
The column \emph{Win(A)} reports the proportion of simulations in which option~A prevailed, while $\overline{t}_{\mathrm{cons}}$ gives the mean time to consensus in simulation steps. 
This design allows us to assess whether increased arousal alone can influence decision dominance and consensus speed when valence is not a differentiating factor.

The outcome indicates that higher $a_A$ does not necessarily lead to dominance at all times. However, configurations with low-to-moderate $v_A$ tend to have a greater win likelihood for~A, indicating that higher arousal itself may give an edge competitively even in the absence of a valence difference. Consensus times show mixed trends, with some high-arousal configurations reaching consensus faster, particularly when $v_A$ is low, possibly reflecting more aggressive recruitment dynamics.
These patterns in general reinforce the conclusion that arousal tends to push decision outcomes one way, though modulated by the valence prevailing between both options in the background.

Figure~\ref{fig:sc5} illustrates the results for scenario~3, where both options initially have equal numerical support and identical emotional states. 
Despite this symmetry, stochastic fluctuations combined with recruitment dynamics lead to the emergence of a dominant option over time. 
The observed curve shows a slow initial growth in the maximum commitment level, followed by a faster ascent once an intermediate support threshold is crossed, consistent with the hypothesised \emph{snowball effect}. 
This behaviour is important because it demonstrates how collective decisions can shift abruptly once a tipping point is reached, even in the absence of initial advantages in size or emotional appeal. 
Such non-linear amplification effects are relevant for understanding real-world phenomena in which minor, random imbalances—once amplified—drive rapid consensus, as seen in both biological swarms and human opinion dynamics.

\section{Analysis of Results}
\label{sec:analysis}

The three simulation scenarios offer complementary insights into how emotional modulation, here represented by valence and arousal, interacts with the classical recruitment--inhibition framework for collective decision-making, as described in \cite{pais2013mechanism,Chase25}. 
While the original bee-equation treats the recruitment rate $r$ and the cross-inhibition rate $\sigma$ as fixed parameters, in our model these rates are effectively modulated by the emotional states of agents. 
High valence can be interpreted as increasing the attractiveness of an option, analogous to raising its effective recruitment rate. In contrast, high arousal can be viewed as amplifying the frequency or intensity of recruitment and inhibition interactions. 

\subsection{Emotional Modulation and Classical Dynamics}
In the original formulation, the balance between $r$ and $\sigma$ determines both the speed of convergence and the probability that a given option prevails. 
Our manipulations of valence and arousal act as multiplicative factors on these rates, meaning that differences in emotional states shift the decision dynamics without altering the underlying structural rules. 
This interpretation aligns with prior theoretical work showing that stronger recruitment (higher effective $r$) accelerates consensus and biases the outcome, while stronger inhibition (higher effective $\sigma$) can slow convergence or prevent deadlock \cite{pais2013mechanism}.

\subsection{Scenario-by-Scenario Interpretation}
Scenario~1 (\emph{Joint Valence--Arousal Influence}) demonstrates that simultaneous increases in valence and arousal for one option produce both higher win rates and shorter consensus times. 
In terms of the classical model, both $r$ and $\sigma$ are effectively increased for the favored option, leading to faster commitment accumulation and more effective suppression of the competitor. 
The non-linear interaction observed between valence and arousal echoes the recruitment--inhibition balance: boosting one rate in isolation has a moderate effect, but increasing both together can dramatically alter the system trajectory \cite{Chase25}.

Scenario~2 (\emph{Arousal Tie-Break}) isolates the role of arousal when valence is held constant. 
Here, differences in arousal map primarily onto differences in the effective rate of recruitment interactions, with the inhibition rate remaining symmetrical. 
The results indicate that higher arousal can still bias the outcome, particularly at low to moderate valence levels, by increasing the likelihood that uncommitted agents encounter and join the high-arousal group. 
This effect aligns with the original model's prediction that recruitment asymmetries, even without differences in attractiveness, can break ties and drive consensus \cite{pais2013mechanism}.

Scenario~3 (\emph{Snowball Effect}) begins with perfect symmetry in both population distribution and emotional state, so $r$ and $\sigma$ are identical across options at $t=0$. 
In this case, the emergence of a winner is driven purely by stochastic fluctuations and the non-linear amplification inherent in the recruitment process. 
Once a random imbalance pushes one option above an intermediate threshold, the self-reinforcing loop of recruitment accelerates commitment growth, leading to rapid dominance, a phenomenon explicitly identified as a tipping point or symmetry-breaking event in \cite{pais2013mechanism,Chase25}.

\subsection{Emotional Dynamics}
From an intuitive standpoint, valence acts like the \emph{appeal} of an option; the more pleasant or attractive it seems, the easier it is to convince others to join. 
Arousal, on the other hand, is the \emph{energy} or urgency with which supporters recruit others. 
Scenario~1 shows that having both appeal and energy on your side is the fastest path to victory; scenario~2 shows that energy alone can still tip the balance when neither side has more appeal; and scenario~3 shows that even without emotional differences, small random leads can snowball into overwhelming consensus.

\subsection{Connecting the Scenarios}
Together, the three scenarios highlight two key mechanisms behind collective decisions:
\begin{enumerate}
    \item \textbf{Emotional modulation of structural parameters}: Changes in valence and arousal effectively shift the recruitment and inhibition rates of the classical bee-equation, producing predictable biases and convergence speeds \cite{pais2013mechanism}.
    \item \textbf{Intrinsic non-linear amplification}: Even in the absence of emotional asymmetry, the recruitment process can amplify small advantages into decisive wins once a tipping point is reached \cite{Chase25}.
\end{enumerate}
These findings bridge biological swarm theory with socio-emotional modelling, suggesting that both emotional states and intrinsic recruitment dynamics must be considered when predicting or influencing the outcome of collective decisions.

\section{Conclusions}
\label{sec:conclusions}

This study introduced an extension of the classical bee-equation framework by incorporating emotional parameters (valence and arousal) into the dynamics of recruitment and inhibition. 
While the original model operates with fixed rates, our approach treats these rates as flexible and sensitive to the agents' affective state. 
This design not only broadens the biological inspiration of the model but also makes it applicable to a broader range of collective systems in which emotions, attitudes, or motivational states play a role in shaping decisions.

Beyond reproducing well-known swarm dynamics, our simulations reveal how emotional modulation can serve as a control lever in collective systems. 
By adjusting the emotional profile of individuals, it is possible to bias group outcomes, accelerate or delay consensus, and even influence the stability of competing states.  Such insights could inform strategies in domains ranging from distributed robotics to online community management, where group-level behavior emerges from repeated local interactions.

The addition of emotion-based variability also provides a methodological window into connecting natural and social systems research. From a biological perspective, valence and arousal can be viewed as measures of inner motivational drives or external cues impacting recruitment and inhibition. In everyday situations involving people, they chime with psychological variables like enthusiasm, urgency, or perceived desirability. This dual interpretability allows the model to act as a conceptual bridge, supporting comparative studies across species and decision environments.

Finally, the scenarios explored here illustrate that collective decisions are not only the product of structural dynamics but can also be steered by modulating affective parameters. 
This suggests that interventions aiming to shift group outcomes operate effectively at the emotional or motivational level, rather than solely by altering structural connectivity or information flow. 
Future work could expand this approach by incorporating dynamic emotional feedback, heterogeneous agent sensitivities, or multi-option settings, thereby deepening our understanding of how affect shapes the tempo and trajectory of collective choice.

\section*{\uppercase{Acknowledgements}}


This work is partially funded funded by project PID2021-122402OB-C22/MICIU/AEI
/10.13039/501100011033 FEDER, UE and by the ACIISI-Gobierno de Canarias and European FEDER funds under project ULPGC Facilities Net and Grant \mbox{EIS 2021 04}. 

\bibliographystyle{apalike}

\end{document}